# Hybridization, Inter-Ion Correlation, and Surface States in the Kondo Insulator SmB$_6$

*Xiaohang Zhang,*[1,*] *N. P. Butch,*[2] *P. Syers,*[1] *S. Ziemak,*[1] *Richard L. Greene,*[1] *and J. Paglione*[1]

[1]Center for Nanophysics and Advanced Materials & Department of Physics, University of Maryland, College Park, Maryland 20742

[2]Condensed Matter and Materials Division, Lawrence Livermore National Laboratory, Livermore, California 94550

[*] Present Address: National Institute of Standards and Technology, Gaithersburg, Maryland 20899

**ABSTRACT**. As an exemplary Kondo insulator, SmB$_6$ has been studied for several decades; however, direct evidence for the development of the Kondo coherent state and the evolution of the electronic structure in the material has not been obtained due to the rather complicated electronic and thermal transport behavior. Recently, these open questions attracted increasing attention as the emergence of a time-reversal invariant topological surface state in the Kondo insulator has been suggested. Here, we use point-contact spectroscopy to reveal the temperature dependence of the electronic states in SmB$_6$. We demonstrate that SmB$_6$ is a model Kondo insulator: below 100 K, the conductance spectra reflect the Kondo hybridization of Sm ions, but below $\sim$ 30 K, signatures of inter-ion correlation effects clearly emerge. Moreover, we find evidence that the low-temperature insulating state of this exemplary Kondo lattice compound harbors conduction states on the surface, in support of predictions of nontrivial topology in Kondo insulators.

Subject Areas: Condensed Matter Physics, Strongly Correlated Materials, Topological Insulators.

The many-body screening interaction between a single localized spin and a continuum of conduction electrons known as the Kondo effect is a well-established element of many physical systems, such as noble metals [1], quantum dots [2], and graphene [3], etc. When spins are arranged in a periodic array, or Kondo lattice, the hybridization leads further to the characteristic opening of an energy gap in the electronic density of states [4]. However, this signature in the electronic structure has not yet been unambiguously observed in Kondo insulators.

Unlike conventional band insulators, a Kondo insulator (KI) features an energy gap in the electronic density of states (DOS) whose magnitude is strongly temperature-dependent and only fully developed at low temperatures. This sensitivity to temperature is rooted in the electronic interactions underlying the Kondo effect. At high temperatures, localized spins on ions are only weakly coupled to the conduction electrons in their host material. As the temperature is lowered below a material-dependent characteristic Kondo temperature, the localized spin states hybridize with the itinerant electrons, forming a many-body spin singlet state. When considering only a single isolated ion, the preceding description is enough, but in a Kondo lattice, the proximity of periodic ions brings into play additional correlations upon further cooling, resulting in the reconstruction of the electronic band structure and the formation of a hybridization gap in the DOS [4]. The KI is a special case of the Kondo lattice, in which the Fermi level or chemical potential falls in the gap (FIG. 1b), in contrast to heavy fermion metals [5], in which the Fermi level coincides with a finite (and large) DOS.

Although this phenomenology is widely accepted, the experimental evidence for the temperature evolution of the electronic states in these complex systems is incomplete. An ideal tool to address the issue is quasiparticle tunneling spectroscopy, which directly probes the DOS. In the more straightforward case of a single-ion Kondo system, the conductance spectrum reflects the interference between the two paths that an injected electron can take (FIG. 1a): one directly to the itinerant electrons and the other indirectly through the hybridized many-body Kondo state. This interference causes a characteristic Fano resonance line shape [6], as first observed in scanning tunneling spectroscopy (STS) measurements on single magnetic adatoms embedded in normal metals [7].

In a Kondo lattice, the development of the correlated ground state should substantially modify the Fano lineshape [8-18]. However, these effects have not been clearly observed in STS [13-16] and point-contact spectroscopy (PCS) [17,18] measurements on heavy-fermion metals. It is possible that signatures of correlation effects may be clouded by the presence of competing interactions underlying tendencies towards superconductivity, magnetism, or quantum criticality [4,5]. For example, substantial differences in the detailed characteristics and the temperature dependence of the conductance spectra obtained on $URu_2Si_2$ [13,14,18] hold different implications for the relationship between the Kondo ground state and the mysterious hidden order phase. In contrast, Kondo insulators, which lack competing ordered phases, make an attractive test bed for the study of electronic interactions in Kondo lattices.

One such material is $SmB_6$, which crystallizes as a simple cubic lattice of samarium ions separated by boron octahedra (inset of FIG. 1a). The basic ingredients of local-itinerant hybridization have long been recognized: the electronic configuration of the Sm ions fluctuates between $4f^6$ ($Sm^{2+}$) and $4f^55d$ ($Sm^{3+}$), giving an effective valence of 2.5 at low temperatures [19,20]; there is a 20 meV gap in the optical conductivity [21]; the low temperature magnetic susceptibility is indicative of a nonmagnetic ground state [22]; the temperature dependence of the electrical resistivity is insulating, increasing by several orders of magnitude upon cooling [23], while the carrier density decreases to $10^{17}$ cm$^{-3}$ [24]. Yet, the validity of the Kondo insulator description for $SmB_6$ is not agreed upon [23,25-28]. One point of particular concern has been that there is finite electrical resistivity at very low temperatures, whereas in a fully gapped insulator it should be infinite [25-27]. Our new measurements clarify this issue, showing that $SmB_6$ acts like a model Kondo insulator [8], and the absence of true insulating behavior at low temperatures reflects the influence of surface conducting states [29].

Point-contact spectroscopy is a versatile tool for the study of the electronic structure in various materials, including superconductors [30,31] and ferromagnets [32,33]. It has been demonstrated that the point-contact spectroscopy technique can be used to probe the bulk DOS and study the many-body interactions in strongly correlated materials [18, 34, also see Part I of the Supplementary Information]. Previous measurements on $SmB_6$ [35-38] have reported an asymmetric line shape with gap-like features in the low-temperature point-contact spectroscopy, but a definitive explanation of the shape and temperature evolution are lacking. In our study, the temperature dependence of the differential point-contact conductance data, $dI/dV$, measured using a $Ag/SmB_6$ junction is summarized in FIG. 1c. Below 100 K, the zero-bias conductance starts to decrease, forming a trough with a half-width of about 20 mV at low temperatures (FIG. 2a). The emergence of this feature reflects the onset of Kondo hybridization between the Sm local moments and the conduction electrons and is consistent with other properties. At a similar temperature, the magnetic susceptibility deviates from high-temperature independent-spin paramagnetism, leading to a broad hump at lower temperatures [22]. X-ray absorption spectroscopy studies also show a substantial decrease in Sm valence starting at about 100 K [19]. In addition, angle-resolved photoemission measurements indicate that a gap in the DOS opens at about 100 K [28]. As temperature decreases, the point-contact conductance spectrum develops an asymmetry that becomes especially pronounced below 40 K, where a conductance peak clearly appears at 20 mV. Although the relationship between similar spectroscopy features and Kondo hybridization was previously questioned [37], upon quantitative examination of our conductance spectra, it becomes clear that the asymmetric peak-trough structure matches theoretical predictions for tunneling into Kondo lattices [8-12].

Beginning with the classical Fano spectral line shape [6], the differential conductance is given by

$$G(V) \propto \frac{((eV - E_0)/\Gamma + q)^2}{1 + ((eV - E_0)/\Gamma)^2} \quad (1)$$

where $E_0$ is the Kondo resonance energy, $\Gamma$ is the full width at half maximum, and $q$ is the Fano factor, which reflects the ratio of probabilities between the two tunneling paths. The best Fano resonance fits (Supplementary information) are presented as dashed curves in FIG. 2a. Clearly, the classical Fano model provides a quantitative description for the development of the zero-bias dip as well as the asymmetric conductance lineshape as temperature decreases. In particular, the simulations well captured the conductance spectra at temperatures in a broad range from 100 K to 30 K. Moreover, the temperature dependence of the fitting parameters (Supplementary Information) establishes a phenomenological understanding for the asymmetric point-contact spectroscopy in the framework of Kondo resonance. Despite the great success in applying the classical Fano model, we noticed that the original Fano resonance lineshape fails to precisely describe the conductance spectra below 30 K, or once the asymmetric peak structure at 20 mV becomes evident.

It is important to note that the classical Fano line shape (Eq. 1) explains the tunneling spectrum of single magnetic impurity systems, in which the DOS is temperature independent and the thermal evolution of the spectrum is solely attributed to the Kondo resonance. In contrast, the local moments in Kondo lattices are strongly coupled to each other; and the collective interactions result in a reconstruction of the electronic DOS that must be simultaneously considered. Indeed, recent studies on heavy-fermion metals [15,18] have also suggested the inadequacy of the classical Fano analysis.

To account for the evolution of the electronic structure in Kondo lattices, Maltseva *et al.* [8] calculated the tunneling current based on the large-*N* mean field theory. The differential tunneling conductance obtained in the study is given by

$$G(V) \propto \operatorname{Im} \tilde{G}_\psi^{KL}(eV); \quad \tilde{G}_\psi^{KL}(eV) = (1 + \frac{q\Delta}{eV - \lambda})^2 \ln\left[\frac{eV + D_1 - \frac{\Phi^2}{eV - \lambda}}{eV - D_2 - \frac{\Phi^2}{eV - \lambda}}\right] + \frac{2D/t_C^2}{eV - \lambda} \quad (2)$$

where $-D_1$ and $D_2$ are respectively the lower and the upper conduction band edges, $2D = D_1 + D_2$ is the bandwidth, $\Delta$ is the width of the Kondo resonance, $\lambda$ is the renormalized *f*-level and $2\Phi$ is the direct gap in momentum space. The Fano factor $q = (\Phi t_f)/(\Delta t_c)$, where $t_f$ and $t_c$ are the matrix amplitudes for tunneling into the Kondo states and into the itinerant electrons, respectively. Finally, the indirect hybridization gap $\Delta_{hyb.}$ is given by $2\Phi^2/D$. To represent the quasiparticle broadening effect, a $\gamma_0$ term is introduced [39]. As a consequence, the energy, *eV*, in Eq. 2 is replaced by $eV - i\gamma_0$. With a background representing the DOS at high temperatures, the best fits (Supplementary information) to our experimental data based on Eq. 2 are shown as solid lines in FIG. 2a. The results indicate a nearly constant $\lambda$ term of ~0.5 meV in the whole temperature range (FIG. 2b), suggesting that the renormalized *f*-level is pinned slightly above the Fermi-level of the Kondo insulator. The simulations also result in a constant Kondo resonance width of about 6.7 mV, which is consistent with the scale of the Kondo temperature (~80 K). FIG. 2e shows the temperature dependence of the hybridization gap calculated from the hybridization amplitude (FIG. 2d) and the band width. At temperatures below 30 K, the hybridization gap retains a constant value of about 18 meV, which is consistent with the gap size determined by low-temperature resistivity [23,40], specific heat [41], and optical reflectivity studies [21]. The hybridization gap decreases at higher temperatures and completely vanishes at a temperature around 100 K.

The remarkable agreement between our experimental data and the model of Maltseva *et al.* (Eq. 2) [8] provides the first direct evidence for the simultaneous development of the Kondo coherent state and the hybridization gap in a Kondo insulator. Based on the temperature range (30-40 K) where the classical Fano lineshape fails to adequately describe the data, we conclude that correlation effects and the ensuing change in the DOS play a physically significant role at surprisingly high temperatures. In fact, in this temperature range, the electrical resistivity increases dramatically and the Hall voltage changes sign [24], which links the telltale characteristics of $SmB_6$ convincingly to Kondo correlation effects.

One unaddressed question remains, however, and it is central to the controversy surrounding the proper description of $SmB_6$: why does the resistivity not diverge, but become temperature-independent below 4 K (FIG. 3a)? The answer has typically been that it is due to in-gap states [42], either due to impurities, incomplete hybridization, or some more exotic scenario. There is surprisingly little direct evidence that any in-gap bound states exist – claims of a 3.5 meV activation energy are based on fits to resistivity data from 5-50 K, specifically in a temperature range where the hybridization gap is temperature dependent and activated behavior would not be expected.

A solution to this conundrum was suggested recently by Dzero *et al*. in their groundbreaking proposal of topological Kondo insulators [29], where $SmB_6$ was singled out as the best candidate for realizing a strong topological insulator. It is widely believed that the unusual residual resistivity results from the presence of a metallic in-gap conduction channel at low temperatures, and candidate states have been proposed to exist in the bulk or on the surface (FIG. 3b and 3c). Our spectroscopic measurements of the bulk Kondo insulator state allow us to determine which of these scenarios is correct. If the saturation of the bulk resistivity were due to the coherent transport of bulk in-gap states [20,40,42], the electronic structure would change at ~ 4 K as a coherent conduction channel develops. In terms of spectroscopy, the enhanced DOS at the Fermi level would lead to a zero-bias peak in the point-contact spectroscopy measurement, or at the very least excess spectral weight. Instead, our results suggest that the bulk DOS is stable and constant in the entire temperature range below 10 K. Thus we rule out the presence of bulk in-gap states. It is clear then that the temperature dependence of the electrical resistivity must reflect a competition between the true Kondo insulator in the bulk and a metallic surface state. Within this picture, the resistivity of $SmB_6$ crystals reflects the activated behavior due to the presence of the hybridization gap in the bulk [20,42] except at the lowest temperatures where surface conduction becomes predominant. As shown in FIG. 3d, a simple two-channel transport simulation based on this picture clearly indicates that the bulk resistivity approaches a saturation point at ~4 K. Despite the success of this explanation, we note that the topological character of the surface metallic states has yet to be determined. For example, the presence of topologically trivial surface electronic states might arise from electronic or structural surface reconstruction, as possibly exemplified by a higher $Sm^{3+}/Sm^{2+}$ ratio at the surface of $SmB_6$ [43]. To address this critical question, surface-dominant tunneling spectroscopic measurements are currently being pursued.

During the review of this manuscript, two preprints posted on arXiv [44,45] have given transport evidence for a conducting surface state in $SmB_6$ at low temperatures. This is a direct confirmation of the conclusions of our point-contact spectroscopy studies of $SmB_6$ reported here.

The authors acknowledge fruitful discussions with A. Nicolaou, J. Zhou, G. Levy, and A. Damascelli at University of British Columbia, V. Galitski at University of Maryland, and M. Dzero at Kent State University. We especially thank Laura H. Greene of University of Illinois at Urbana-Champaign for her thoughtful comments and suggestions. X. Z. and N. P. B. would also like to thank K. Jin, S. R. Saha and R. W. Hu for technical assistance. The work at University of Maryland was supported in part by the NSF under Grant Nos. DMR-1104256 and DMR-0952716, and the AFOSR-MURI under Grant No. FA9550-09-1-0603.

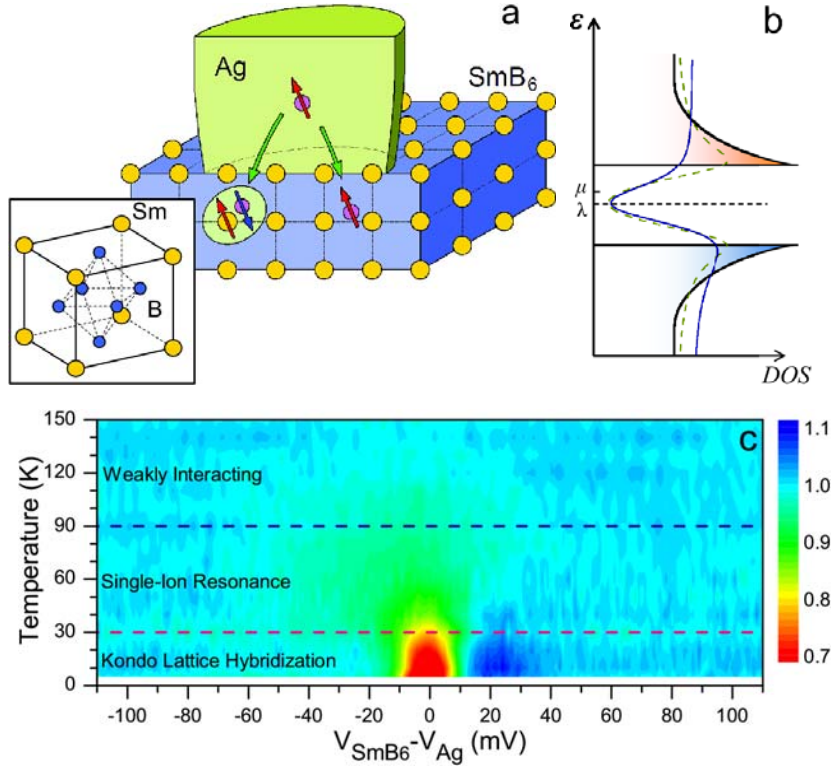

FIG. 1. (a) A schematic view for a Ag/Sm$_6$ junction. A Fano-like conductance spectrum is expected due to the interference of the two electron paths as indicated by two green arrows: one is directly into the itinerant electron states (right side), thus should reflect the DOS; the other is indirectly into the electron sea through a spin-singlet Kondo state (left side). Inset: the crystal structure of SmB$_6$ single crystal. The Sm ions form a simple cubic structure. (b) A schematic view of the DOS for the ground state of SmB$_6$. In point-contact spectroscopy measurement, quasiparticle broadening should be taken into account as indicated by the green dashed line. The blue solid curve is the expected spectrum with Fano resonance. μ is the chemical potential while λ is the renormalized *f*-level. (c) The contour map of the normalized conductance in bias voltage-temperature (*V-T*) plane. All conductance data were normalized by the spectrum conductance obtained at 150 K. As indicated by the color label, red and blue colors are used to represent values below 0.7 and above 1.1, respectively. At temperatures above ~ 90 K, the conductance spectroscopy of the junction is highly symmetric with negligible temperature dependence due to the weak interaction between electrons and local moments. From 90 K to 30 K, the zero-bias conductance is gradually suppressed while an asymmetric conductance lineshape develops. In this temperature range, the conductance spectrum can be described by the single-ion resonance model. Below ~ 30 K, the conductance spectra are clearly asymmetric with a dip at zero bias and a peak at 20 mV. As discussed in the main text, this strongly asymmetric conductance lineshape provides a direct evidence for the inter-ion correlations in the Kondo lattice system.

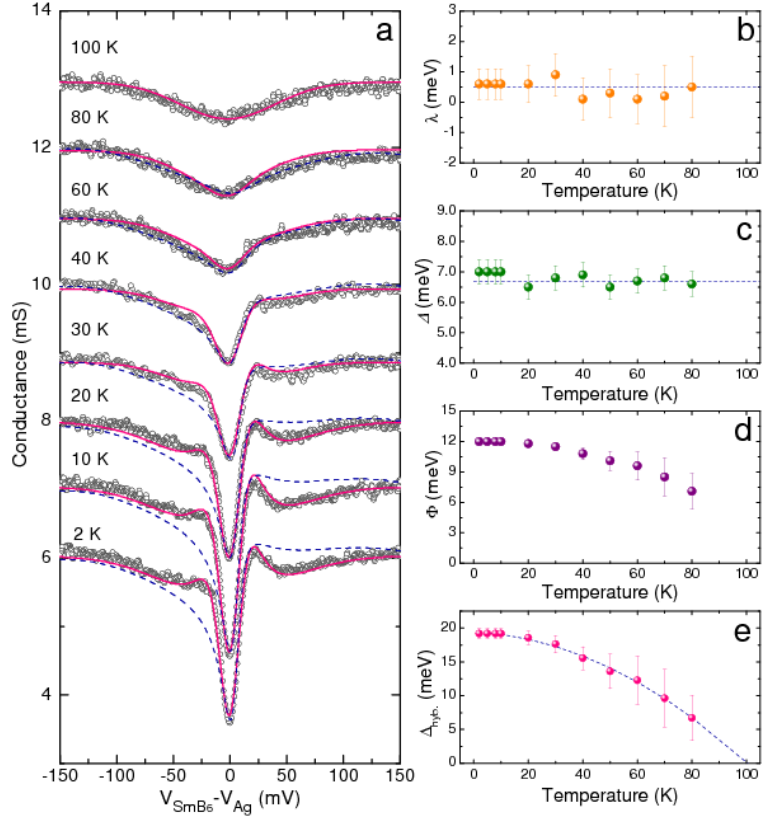

FIG. 2. (a) The observed conductance spectra with the best fits to the classical Fano resonance line shape (Eq. 1 and dashed lines) and to the tunneling model for Kondo lattices (Eq. 2 and solid lines), respectively. For clarity, only the spectrum and the simulations for 2 K are plotted with the actual value, while other curves are vertically shifted; (b) (c) and (d) show the temperature dependence of λ (renormalized *f*-level), Δ (Kondo resonance width), and Φ (half of the direct gap size), respectively used in the Kondo lattice tunneling simulations. The dashed line in (b) and (c) suggest that both the renormalized *f*-level and the Kondo gap width are nearly temperature independent in the entire temperature range below 100 K; (e) The temperature dependence of the hybridization gap extracted from the hybridization amplitude and the band width D based on the Kondo lattice tunneling model. The dashed line in (e) represents a smooth trend of the size of the hybridization gap as the temperature changes. The extrapolation of the curve suggests that the hybridization gap vanishes at a temperature around 100 K.

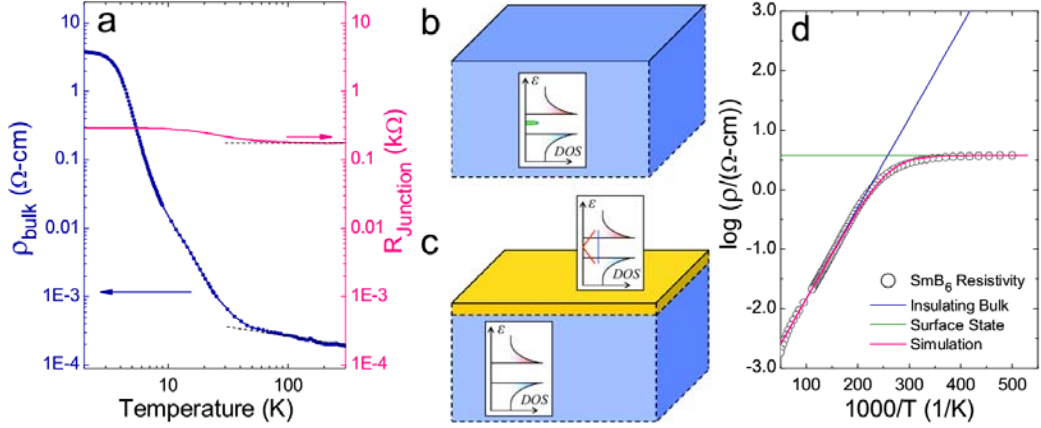

FIG. 3. (a) Temperature dependence of electrical resistivity of a $SmB_6$ single crystal (blue) and the resistance of a $Ag/SmB_6$ junction (pink). Compared to the four-order of magnitude increase in the resistivity and a significant suppression below 4 K for $SmB_6$ single crystals, the temperature dependence of the $Ag/SmB_6$ junction resistance exhibits a rather different behavior with an overall increase of about 2 in the measured temperature range and a low-temperature saturation below ~ 10 K. The dashed lines indicate that the junction resistance holds a constant value at temperatures above ~ 100 K while the resistivity of bulk single-crystals shows an activation behavior in the same temperature range. (b) Schematic view of the DOS in $SmB_6$ without the presence of a surface state. It is expected that the coherent transport of the bulk in-gap bound states (green area) leads to the suppression in resistivity at temperatures below 4 K; (c) The DOS in $SmB_6$ with the emergence of a surface state at low temperature. The finite residue resistivity of the material is due to the presence of a conducting surface state. Both topological surface states (orange) and conventional two-dimension electron gas (blue) could lead to such a metallic 2D conduction; (d) Simulation of the measured resistivity in the scenario that a metallic-like surface state dominates the conductance at low temperatures. The overall conductivity is the sum of two conduction channels: the surface conduction (green) and the intrinsic bulk conduction (blue). The measured resistivity approaches the saturation point at a temperature around 4 K.
8

REFERENCES

[1]  J. Kondo, *Resistance minimum in diluted magnetic alloys*, Prog. Theor. Phys. **32**, 37-49 (1964).

[2]  D. Goldhaber-Gordon, H. Shtrikman, D. Mahalu, D. Abusch-Magder, U. Meirav, and M. A. Kastner, *Kondo effect in a single-electron transistor*, Nature **391**, 156 (1998)

[3]  J.-H. Chen, L. Li, W. G. Cullen, E. D., Williams and M. S. Fuhrer, *Tunable Kondo effect in graphene with defects*, Nature Phys. **7**, 535 (2011)

[4]  See review by P. Coleman, *Fundamental Theory*, edited by H. Kronmüller and S. S. P. Parkin, Handbook of Magnetism and Advanced Magnetic Materials Vol. 1 95-148 (John Wiley and Sons, New York, 2007)

[5]  A.C. Hewson, The Kondo Problem to Heavy Fermions (Cambridge University Press, Cambridge, England, 1993)

[6]  U. Fano, *Effects of configuration interaction on intensities and phase shifts*, Phys. Rev. **124**, 1866 (1961)

[7]  For example: V. Madhavan, W. Chen, T. Jamneala, M.F. Crommie, and N.S. Wingreen, *Tunneling into a single magnetic atom: spectroscopic evidence of the Kondo resonance*, Science **280**, 5363 (1998)

[8]  M. Maltseva, M. Dzero, and P. Coleman, *Electron cotunneling into a Kondo lattice*, Phys. Rev. Lett. **103**, 206402 (2009)

[9]  Y.-F. Yang, *Fano effect in the point contact spectroscopy of heavy-electron materials*, Phys. Rev. B **79**, 241107(R) (2009)

[10] J. Figgins and D.K. Morr, *Differential conductance and quantum interference in Kondo systems*, Phys. Rev. Lett. **104**, 187202 (2010)

[11] P. Wölfle, Y. Dubi, and A.V. Balatsky, *Tunneling into clean heavy fermion compounds: origin of the Fano line shape*, Phys. Rev. Lett. **105**, 246401 (2010)

[12] J. X. Zhu, J. P. Julien, Y. Dubi, and A. V. Balatsky, *Local electronic structure and Fano interference in tunneling into a Kondo hole system*, Phys. Rev. Lett. **108**, 186401 (2012)

[13] A.R. Schmidt, M.H. Hamidian, P. Wahl, F. Meier, A.V. Balatsky, J.D. Garrett, T.J. Williams, G.M. Luke, and J.C. Davis, *Imaging the Fano lattice to 'hidden order' transition in $URu_2Si_2$*, Nature **465**, 570 (2010)

[14] P. Aynajian, E.H. da Silva Neto, C.V. Parker, Y. Huang, A. Pasupathy, J. Mydosh, and A. Yazdani, *Visualizing the formation of the Kondo lattice and the hidden order in $URu_2Si_2$*, Proc. Natl. Acad. Sci. USA **107**, 10383 (2010)

[15] S. Ernst, S. Kirchner, C. Krellner, C. Geibel, G. Zwicknagl, F. Steglich, and S. Wirth, *Emerging local Kondo screening and spatial coherence in the heavy-fermion metal $YbRh_2Si_2$*, Nature **474**, 362 (2011)

[16] P. Aynajian, E.H. da Silva Neto, A. Gyenis, R. E. Baumbach, J. D. Thompson, Z. Fisk, E. D. Bauer, and A. Yazdani, *Visualizing heavy fermions emerging in a quantum critical Kondo lattice*, Nature **486**, 201 (2012)

[17] W.K. Park, J.L. Sarrao, J.D. Thompson, and L.H. Greene, *Andreev reflection in heavy-fermion superconductors and order parameter symmetry in $CeCoIn_5$*, Phys. Rev. Lett. **100**, 177001 (2008)

[18] W.K. Park, P.H. Tobash, E. Ronning, E.D. Bauer, J.L. Sarrao, J.D. Thompson, and L.H. Greene, *Observation of the hybridization gap and Fano resonance in the Kondo lattice $URu_2Si_2$*, Phys. Rev. Lett. **108**, 246403 (2012)

[19] M. Mizumaki, S. Tsutsui, and F. Iga, *Temperature dependence of Sm valence in $SmB_6$ studied by X-ray absorption spectroscopy*. J. Phys.: Conf. Ser. **176**, 012034 (2009)

[20] T. Caldwell, A.P. Reyes, W.G. Moulton, P.L. Kuhns, M. J. R. Hoch, P. Schlottmann, and Z. Fisk, *High-field suppression of in-gap states in the Kondo insulator $SmB_6$*, Phys. Rev. B **75**, 075106 (2007)

[21] B. Gorshunov, N. Sluchanko, A. Volkov, M. Dressel, G. Knebel, A. Loidl, and S. Kunii, *Low-energy electrodynamics of $SmB_6$*. Phys. Rev. B **59**, 1808 (1999)

[22] S. Gabáni, K. Flachbart, V. Pavlík, T. Herrmannsdörfer, E.S. Konovalova, Yu. Paderno, J. Bednarčin, and J. Trpčevská, *Magnetic properties of $SmB_6$ and $Sm_{1-x}La_xB_6$ solid solutions*. Czech. J. Phys. **52**, A225 (2002)

[23] J.C. Cooley, M. C. Aronson, Z. Fisk, and P.C. Canfield, *$SmB_6$: Kondo insulator or exotic metal?* Phys. Rev. Lett. **74**, 1629 (1995)

[24] J. W. Allen, B. Batlogg, and F. Wachter, *Large low-temperature Hall effect and resistivity in mixed-valent $SmB_6$*, Phys. Rev. B **20**, 4807 (1979)

[25] K. A. Kikoin and A.S. Mishchenko, *Magnetic excitations in intermediate-valence semiconductors with a singlet ground state*, J. Phys.: Condens. Matter **7**, 307(1995)

[26] B. R. Coles, *Speculations concerning $SmB_6$*, Physica B **230-232**, 718 (1997)

[27] T. Kasuya, *Physical Mechanism in Kondo Insulator*, J. Phys. Soc. Jpn. **65**, 2548 (1997)

[28] S. Nozawa, T. Tsukamoto, K. Kanai, T. Haruna, S. Shin, and S. Kunii, *Ultrahigh-resolution and angle-resolved photoemission study of $SmB_6$*, J. Phys. Chem. Solids **63**, 1223 (2002)

[29] M. Dzero, K. Sun, V. Galitski, and P. Coleman, *Topological Kondo insulators*, Phys. Rev. Lett. **104**, 106408 (2010)

[30] A. Biswas, P. Fournier, M. M. Qazilbash, V. N. Smolyaninova, H. Balci, and R. L. Greene, *Evidence of a d- to s-Wave Pairing Symmetry Transition in the Electron-Doped Cuprate Superconductor $Pr_{2-x}Ce_xCuO_4$*, Phys. Rev. Lett. **88**, 207004 (2002)

[31] X. H. Zhang, Y. S. Oh, Y. Liu, L. Q. Yan, S. R. Saha, N. P. Butch, K. Kirshenbaum, K. H. Kim, J. Paglione, R. L. Greene, and I. Takeuchi, *Evidence of a universal and isotropic $2\Delta/k_BT_C$ ratio in 122-type iron pnictide superconductors over a wide doping range*, Phys. Rev. B **82**, 020515R (2010)

[32] Y. Ji, G. J. Strijkers, F. Y. Yang, C. L. Chien, J. M. Byers, A. Anguelouch, G. Xiao, and A. Gupta, *Determination of the spin polarization of half-metallic $CrO_2$ by point contact Andreev reflection*, Phys. Rev. Lett. **86**, 5585 (2001)

[33] X. H. Zhang, S. von Molnár, Z. Fisk, and P. Xiong, *Spin-dependent electronic states of the ferromagnetic semimetal $EuB_6$*, Phys. Rev. Lett. **100**, 167001 (2008)

[34] H. Z. Arham, C. R. Hunt, W. K. Park, J. Gillett, S. D. Das, S. E. Sebastian, Z. J. Xu, J. S. Wen, Z. W. Lin, Q. Li, G. Gu, A. Thaler, S. Ran, S. L. Bud'ko, P. C. Canfield, D. Y. Chung, M. G. Kanatzidis, and L. H. Greene, *Detaection of orbital fluctuations above the structural transition temperature in the iron pnidtides and chalcogenides*, Phys. Rev. B **85**, 214515 (2012).

[35] G. Güntherodt, W. A. Thompson, F. Holtzberg, and Z. Fisk, *Electron tunneling into intermediate-valence materials*. Phys. Rev. Lett. **49**, 1030 (1982)

[36] I. Frankowski and P. Wachter, *Point-contact spectroscopy on $SmB_6$, $TmSe$, $LaB_6$ and $LaSe$*. Solid State Commun. **41**, 577 (1982)

[37] B. Amsler, Z. Fisk, J. L. Sarrao, S. von Molnár, M. W. Meisel, and F. Sharifi, *Electron-tunneling studies of the hexaboride materials $SmB_6$, $EuB_6$, $CeB_6$, and $SrB_6$*, Phys. Rev. B **57**, 8747 (1998)

[38] K. Flachbart, K. Gloos, E. Konovalova, Y. Paderno, M. Reiffers, P. Samuely, and P. Švec, *Energy gap of intermediate valent $SmB_6$ studied by point-contact spectroscopy*. Phys. Rev. B **64**, 085104 (2001)

[39] D. V. Averin and Yu. V. Nazarov, *Virtual electron-diffusion during quantum tunneling of the electric charge*. Phys. Rev. Lett. **65**, 2446 (1990)

[40] N. E. Sluchanko, V. V. Glushkov, S. V. Demishev, A. A. Pronin, A. A. Volkov, M. V. Kondrin, A. K. Savchenko, and S. Kunii, *Low-temperature transport anisotropy and many-body effects in $SmB_6$*, Phys. Rev. B **64**, 153103 (2001)

[41] S. von Molnár, T. Theis, A. Benoit, A. Briggs, J. Flouquet, and J. Ravex, In *Valence Instabilities*, edited by Wachter, P. and Boppart, H., p. 389 (North-Holland, Amsterdam, 1982)

[42] N. E. Sluchanko, V. V. Glushkov, B. P. Gorshunov, S. V. Demishev, M. V. Kondrin, A. A. Pronin, A. A. Volkov, A. K. Savchenko, G. Grüner, Y. Bruynseraede, V. V. Moshchalkov, and S. Kunii, *Intragap states in $SmB_6$*, Phys. Rev. B **61**, 9906 (2000)

[43] M. Aono, R. Nishitani, C. Oshima, T. Tanaka, E. Bannai, and S. Kawai, *$LaB_6$ and $SmB_6$ (001) surfaces studied by angle-resolved XPS, LEED and ISS*, Surface Science **86**, 631 (1979)

[44] S. Wolgast, C. Kurdak, K. Sun, J. W. Allen, D.-J. Kim, and Z. Fisk, *Discovery of the first topological Kondo insulator: samarium hexaboride*, arXiv:1211.5104

[45] J. Botimer, D.-J. Kim, S. Thomas, T. Grant, Z. Fisk, and J. Xia, *Robust surface Hall effect and nonlocal transport in $SmB_6$: indication for an ideal topological insulator*, arXiv:1211.6769

# Hybridization, Inter-Ion Correlation, and Surface States in the Kondo Insulator SmB$_6$

## *Supplementary Information*

*Xiaohang Zhang,[1,*] N. P. Butch,[2] P. Syers,[1] S. Ziemak,[1] Richard L. Greene,[1] and J. Paglione[1]*

[1]CNAM and Department of Physics, University of Maryland, College Park, Maryland 20742

[2]Condensed Matter and Materials Division, Lawrence Livermore National Laboratory, Livermore, California 94550

[*]Present Address: National Institute of Standards and Technology, Gaithersburg, Maryland 20899

## I.   Method Summary

The SmB$_6$ single crystals used in this study were grown by the aluminum-flux method. Soft point-contact junctions were fabricated on fresh single crystals with silver paste as the counterelectrodes. The crystals were rinsed by diluted NaOH solution prior to junction fabrication in order to remove possible residual flux at the surface. The sizes of the Ag/SmB$_6$ contacts were controlled to be about 100 μm in diameter and the measured junction resistance ranged from ~200 Ω to ~2 kΩ. Our junctions showed essentially the same conductance spectroscopy and quantitative analyses presented here are focused on a representative junction. Throughout this work, the bias voltage is defined as the electric potential of the SmB$_6$ crystal with respect to the silver electrode.

Previous experimental work has indicated that the point-contact spectroscopy technique is a powerful probe for bulk electronic properties in strongly correlated materials [18 and 34]. Moreover, we have carefully examined for possible junction interface effects that are not related to the bulk properties of SmB$_6$. First, the conductance spectra measured on our Ag/SmB$_6$ junctions are highly symmetric at temperatures above ~120 K, which rules out the possibility of a Fermi-level mismatch-induced asymmetric behavior in the measurement. Moreover, as shown in FIG. 3a, the zero-bias junction resistance (also see FIG. S2c in linear scale) is nearly temperature independent from 300 K down to about 150 K and only shows an increase of about two in the entire measured temperature range. This temperature dependence is clearly different from that of the bulk resistivity of SmB$_6$ single crystal (also see FIG. S1a). The latter shows an overall increase of four orders of magnitude in the temperature range from 300 K to about 4 K. Therefore, the measured junction resistance is not dominated by the bulk resistance of SmB$_6$ single crystal. Taking all these facts together, we conclude that our point-contact spectroscopy measurements indeed detect the bulk electronic states of $SmB_6$.

## II. Crystal characterization

In this study, $SmB_6$ single crystals grown by the aluminum-flux method [S1] are in rectangular-solid shape with each edge measured typically several hundreds of micrometers. A representative scanning electron micrograph of the crystals is shown in the inset of Figure S1a. The resistivity of the single crystals shows an overall four order of magnitude increase from room temperature to 2 K (Figure S1a). A Curie-Weiss like behavior at temperatures above 100 K is observed in the magnetic susceptibility of the single crystals (Figure S1b). The susceptibility data is suppressed from the Curie-Weiss behavior and shows a broad hump at low temperatures.

## III. Scaling of the peak observed at ~ 20 meV

As a direct evidence for the asymmetric line shape, a peak consistently appears at a positive bias voltage of about 20 meV in low-temperature conductance spectra. To further study the peak as to its origin, we first estimated the height of the peak for each temperature: a fit to the curve obtained at 100 K was subtracted from each conductance spectrum as a background (Figure S2a and S2b) and the peak value was then read from the subtracted spectrum. Here, $h(T)$ was used to denote the height of the peak at temperature $T$. To compare the temperature dependence of the peak at 20 mV with that of the zero-bias junction resistance, we rescaled the peak height as

$$H(T) = h_0 + a \times h(T) \quad \text{(S1)}$$

By choosing proper values respectively for the two constants, $a$ and $h_0$, we were able to demonstrate that the two features show essentially the same temperature dependence (Figure S2c), which strongly suggests that the dip at zero-bias and the peak at 20 mV have the same physical origination.

## IV. Fitting the conductance spectra to the classic Fano formula

To fit the conductance spectra, the conductance spectrum obtained at 100 K was used to describe the background. Based on the classic Fano line shape, the overall conductance spectrum at each temperature is described by

$$G(V) = G_{100K}(V) + w_{CF} G_{CF}(V) \quad \text{(S2)}$$

where $w_{CF}$ is a weight factor, $G_{CF}(V)$ is the classic Fano line shape (Eq. 1 of the main text), while $G_{100K}(V)$ is the background. A fitting example to the classic Fano formula is shown in Figure S3a.

The simulation results for each temperature have been demonstrated in Figure 2a of the main text. Consistent with the temperature dependence of the asymmetric conductance line shape, the Fano factor $q$ (Figure S4a) used in the simulation holds a nearly constant value of ~ 4.5 meV at low temperatures and then gradually diminishes with increasing temperature, indicating the decrease of the coherent proportion. The temperature dependence points to a zero Fano factor at 90-100 K. The Kondo resonance energy $E_0$ (Figure S4b), however, shows no significant temperature dependence in the entire temperature range within the uncertainty of our simulation.

In the single magnetic impurity model, the broadening of the Fano resonance width at finite temperatures is given by [S2]

$$\Gamma = 2\sqrt{(\pi k_B T)^2 + 2(k_B T_K)^2} \qquad (S3)$$

where $k_B$ is the Boltzmann constant. A fit to the extracted low-temperature resonance width (Figure S4c) yielded a Kondo temperature of 75 ± 5 K, consistent with the initial temperature at which the asymmetric conductance lineshape starts to develop.

## V.   Fitting the conductance spectra to the Kondo lattice tunneling model

In the modified Fano model, both the opening of the hybridization gap and the Fano resonance are considered. Again, the conductance spectrum obtained at 100 K was used to describe the background. The conductance spectrum is then given by

$$G(V) = G_{100K}(V) + w_{MF} G_{MF}(V) \qquad (S4)$$

where $w_{MF}$ is a weight factor, $G_{MF}(V)$ is the modified Fano line shape (Eq. 2 of the main text), while $G_{100K}(V)$ is the background. With a background taken at a temperature higher than the Kondo temperature of the system, the variations in the point-contact conductance spectroscopy at lower temperatures are clearly due to the collective effect of the Fano resonance and the opening of the hybridization gap. A fitting example to the modified Fano formula is shown in Figure S3b.

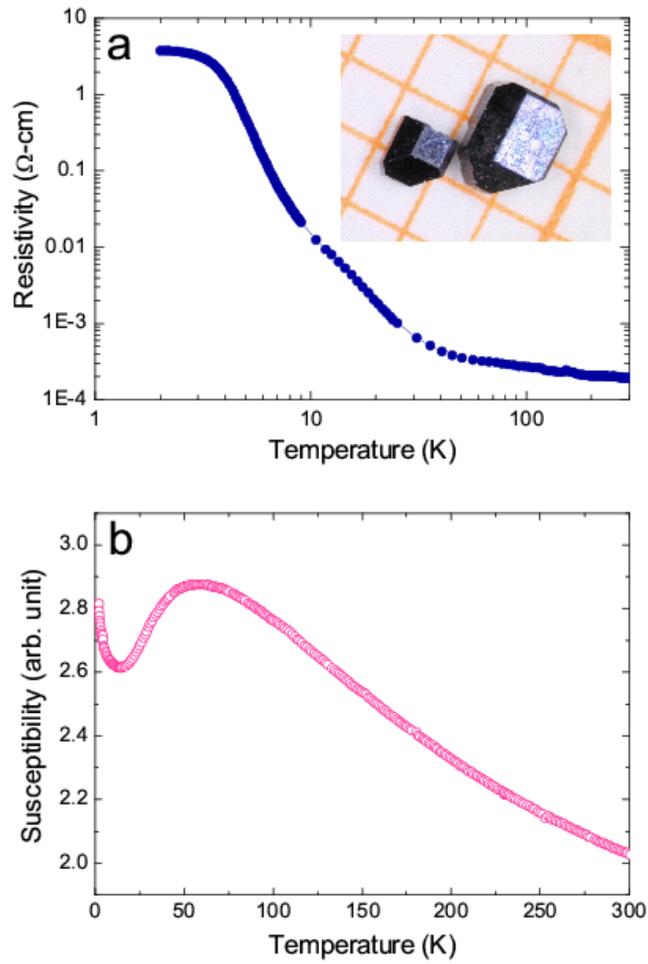

Figure S1. (a) Temperature dependence of the bulk resistivity of $SmB_6$ single crystals. Inset: a representative image of the single crystals. (b) Temperature dependence of the magnetic susceptibility of $SmB_6$ single crystals.

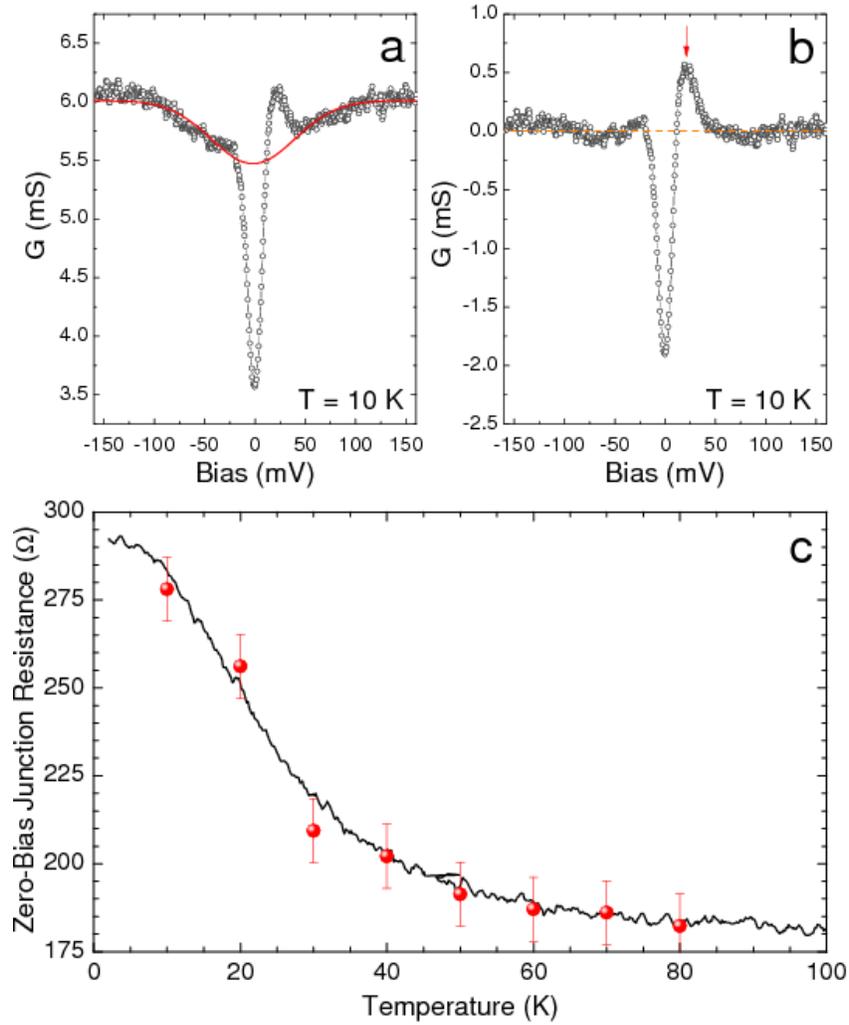

Figure S2. (a) and (b) An example for the estimate of the peak height at around 20 mV: (a) as the background, a fit to the conductance spectrum at 100 K (red line) is subtracted from the raw data (connected black dots); (b) the peak height is directly read from the subtracted spectrum. (c) Scaling of the peak height with the zero-bias junction resistance based on Eq. S1.

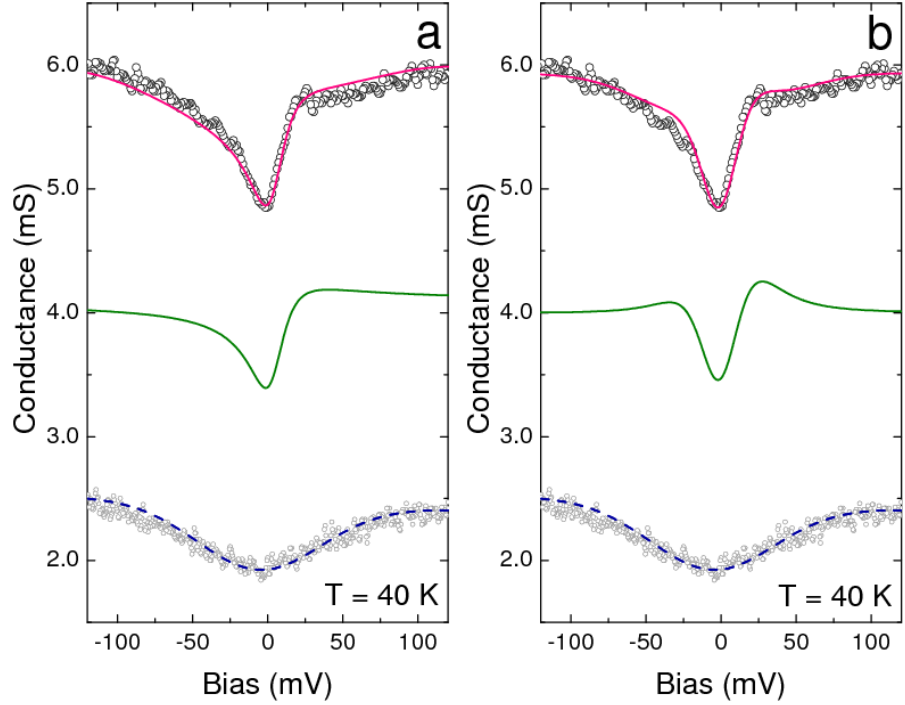

Figure S3. (a) A fitting example based on the classical Fano line shape (Eq. 1 of the main text). The green line is generated from the classic Fano formula with fitting parameters: $q = 0.38$, $E_0 = 4$ meV, and $\Gamma = 28$ meV. The blue dashed line represents the background of the spectrum, which is a fit to the spectrum obtained at 100 K (gray open circles). The pink curve is the fit to the experimental data obtained at 40 K (black open circles). (b) A fitting example based on the Kondo lattice tunneling model (Eq. 2 of the main text). The green line is generated by the Kondo lattice tunneling model with fitting parameters: $D_1 = 12$ mV, $D_2 = 18$ mV, $\Phi = 10.8$ meV, $\Delta = 6.9$ meV, $\lambda = 0.1$ meV, $q = 31$, $t_c = 0.9$, and $\gamma_0 = 29$ meV. Again, the blue dashed line represents the background of the spectrum measured at 100 K. The pink curve is the fit to the experimental data obtained at 40 K (black open circles).

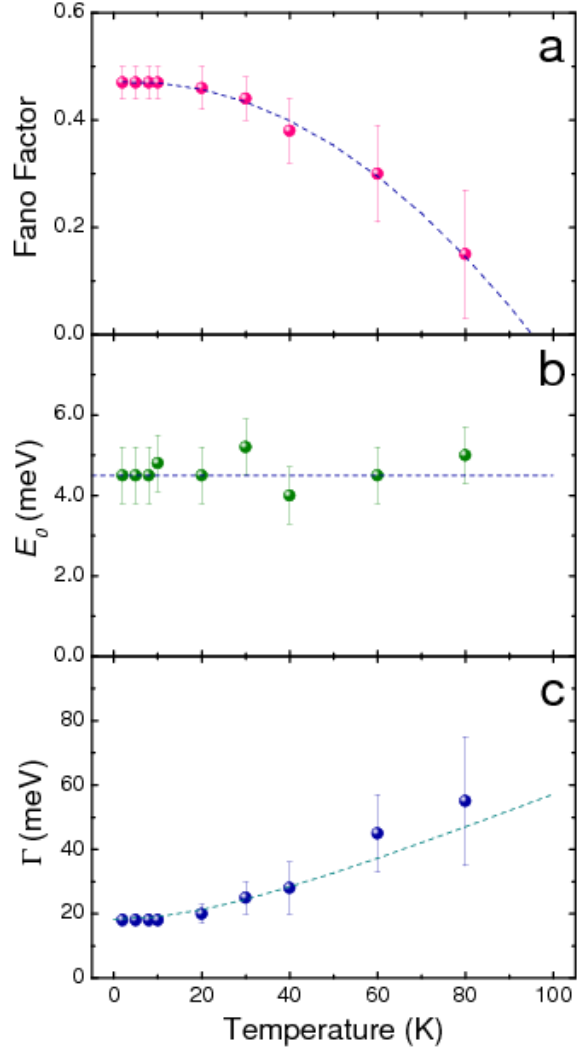

Figure S4. The temperature dependence of the fitting parameters used in the classical Fano resonance simulations (Eq. 1 and dashed curves in Figure 2a of the main text): (a) the Fano factor $q$; (b) the resonance energy $E_0$; and (c) the full width at half maximum (FWHM) $\Gamma$. a smooth trend of Fano factor indicated by the dashed curve in (a) suggests that the Fano resonance vanishes at a temperature around 100 K. The resonance energy remains a nearly constant in the whole temperature range as indicated by the dashed line in (b). The dashed curve in (c) is a fit to Eq. S3 as discussed in the text.

REFERENCES

[S1]  A. Kebede, M. C. Aronson, C. M. Buford, P. C. Canfield, J. H. Cho, B. R. Coles, J. C. Cooley, J. Y. Coulter, Z. Fisk, J. D. Goettee, W. L. Hults, A. Lacerda, T. D. McLendon, P. Tiwari, and J. L. Smith, *Studies of the correlated electron system SmB$_6$*, Physica B **223-224**, 245 (1996)

[S2]  K. Nagaoka, T. Jamneala, M. Grobis, and M.F. Crommie, *Temperature dependence of a single Kondo impurity*, Phys. Rev. Lett. **88**, 077205 (2002)